# Self-induced spontaneous transport of water molecules through a symmetrical nanochannel by ratchetlike mechanism

Rongzheng Wan[1], Hangjun Lu[1,2], Jinyuan Li[1,3], Jingdong Bao[4], Jun Hu[1,5,*], and Haiping Fang[1,*] [1]

[1] Shanghai Institute of Applied Physics, Chinese Academy of Sciences, P.O. Box 800-204, Shanghai 201800, China
[2] Department of Physics, Zhejiang Normal University, 321004, Jinhua, China
[3] Department of Physics, Zhejiang University, Hanzhou, 310027, China
[4] Department of Physics, Beijing Normal University, Beijing 100875, China
[5] Bio-X Life Sciences Research Center, College of Life Science and Technology, Shanghai JiaoTong University, Shanghai 200030, China

**Abstract**

Water molecules, confined in a carbon nanotube, were monitored using molecular dynamics simulation. Spontaneous directional transportation during a long timescale was observed in the symmetrical nanochannel by a ratchet-like mechanism. This ratchet-like system was without any asymmetrical structure or external field, while the asymmetric ratchet-like potential solely resulted from the transported water molecules that formed hydrogen-bonded chains. Remarkably, the resulting net water fluxes reached the level of the biological channel and the average duration for spontaneous directional transportation reached the timescale of many biomolecular functions. This is the first report that heat energy from the surroundings can be used to drive molecules uni-directionally during a long timescale in a nanochannel system. This effect is ascribed to the unique structure of the water molecule.

**Introduction**

Gaining work from thermal fluctuations -- without an input of external energy -- is a dream for scientists, but is forbidden by the second law of thermodynamics [1-3]. Feynman proposed a molecular ratchet in this direction, but theoretical arguments against it remain [1, 2]. In Feynman's molecular ratchet, a "permanent" directional motion without external input energy [1, 2] was attempted. Practically, however, "permanence" may not be necessary and it may only be necessary for the directional motion to be long enough for a certain application. This provides a new direction for designing a practical molecular ratchet. In 1997, Kelly *et al.* designed a molecule containing triptycene with helicenes, and observed spontaneous uni-directional rotations of the triptycene over a short time period [4]. Davis argued that the the second law of thermodynamics survived in this system [5]. Later, spontaneous uni-directional rotations over a short time period have also been observed in many synthetic molecules [6-8]. Those molecules have been further developed as molecular motors, where the thermal fluctuations played important roles [9-15]. Theoretically, Evans *et al.* found that there were considerably large probability for the entropy decreasing for small macroscopic dynamical systems over short time scales although the probability decreases very quickly as the time scales increases and developed the fluctuation theory to characterize those systems [16]. They further proposed that when scaled down to nanometers, there is a significant probability that heat energy from the surroundings can be converted into useful work [17]. Applying this fluctuation theory, Porporato et al. used the relative entropy to quantify time irreversibility [18], and Douarche et al investigated the work fluctuations of a harmonic oscillator in contact with a thermostat and driven out of equilibrium by an external force [19]. In this Letter, this classical problem is revisited using molecular dynamics simulation to monitor water molecules confined in a single-walled carbon nanotube (SWNT). We find that heat energy from the surroundings can be used to drive water molecules uni-directionally during a long timescale in this nanoscale system. We note that water confined in a carbon nanotube usually exhibits unusual properties not seen in bulk [20-30]. When the radius of the SWNT is appropriate, the water molecules form a single file [25-30]. Those single-filed water exhibit conduction in bursts [26], concerted transport that can be well-described by a continuous-time random-walk model [30], bipolar hydrogen orientation orders [31-36]. Very recently, we found that the hydrogen-bond chains in those single-filed water molecules resulted in excellent gating on its mechanical and electrical permeation behavior [37, 38] and controllable pumping ability [39] when there was an appropriate asymmetric charge distribution adjacent to nanochannel. Here, we focus on the observation of spontaneous transport of water molecules across the SWNT without the input of external energy over short time scales. Importantly, the duration for such transport is finite but long, reaching timescales of some biological applications; and the resulting net water fluxes reach the level of biological channel [40].

---

[*] To whom correspondence should be addressed. Email address: jhu@sjtu.edu.cn, fanghaiping@sinap.ac.cn

**Computational Methods**

The SWNT was uncapped [41] with a length of 13.4 Å and diameter of 8.1 Å. To mimic the biological channels in a membrane [31-37], the SWNT was embedded in a graphite sheet along the z-axis. z = 0 corresponds to the left end of the SWNT. The single graphite sheet divided both the full space and the SWNT into two equal parts. The 144-carbon (6, 6) nanotube was formed by folding a graphite sheet of $5 \times 12$ carbon rings to a cylinder, and then relaxing with the interaction between carbon atoms. This interaction has been described by the parameterized potential by Brenner [42], according to the Tersoff formulation [43]. Initially, water molecules are filled in the other space of the system, except for the channel of the SWNT. Periodic boundary conditions are applied in all directions.

Molecular dynamics simulations, which have been used widely for the study of water dynamics in SWNT, proteins, and in-between proteins [26-28, 37, 44-48], were carried out at constant pressure (1 bar with initial box size $L_x = 3.0$ nm, $L_y = 3.0$ nm, $L_z = 4.0$ nm) and temperature (300 K) using the weak coupling scheme of Berendsen with Gromacs 3.2.1. [49], and TIP3P [50] force fields for water. A time step of 2 femtoseconds was used and data were collected every 0.5 picoseconds. In the simulations, the carbon atoms were modeled as uncharged Lennard-Jones particles with a cross-section of $\sigma_{CC} = 0.34$ nanometer (nm), $\sigma_{CO} = 0.3275$ nm, and a depth of the potential well of $\varepsilon_{CC} = 0.3612$ kJ mol$^{-1}$, $\varepsilon_{CO} = 0.4802$ kJ mol$^{-1}$ [26]. Carbon-carbon bond lengths of 0.14 nm and bond angles of 120° were maintained by harmonic potentials with spring constants of 393960 kJ mol$^{-1}$ nm$^{-2}$ and 527 kJ mol$^{-1}$ degree$^{-2}$ before relaxation [49]. In addition, a weak dihedral angle potential was applied to bonded carbon atoms [26].

The SWNT, together with the graphite sheet solvated in a water reservoir, was simulated for 372 nanoseconds (ns) with molecular dynamics, and the last 360 ns were collected for analysis.

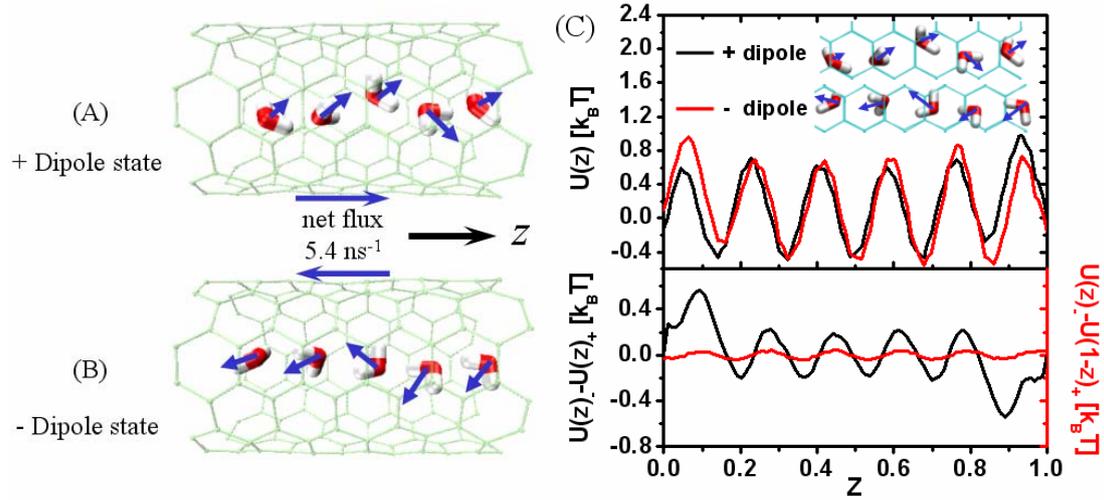

**Figure 1.** Typical cases for (A) + dipole and (B) - dipole states. In + (-) dipole state, all dipoles (blue arrows) of water molecules inside the SWNT are nearly aligned along (opposite to) the $z$ direction. There is an average net flux of 5.4 ns$^{-1}$ of water across the SWNT along the dipole orientation in each state, comparable to that of the biological channel, aquaporin-1 [40]. (C) (Top) The asymmetric *potential of mean force* $U(z)_\pm$ of water molecules inside a SWNT for + and - dipole states; $z$ is the normalized position in the nanotube along the $z$ direction. (Bottom) $U(z)_- - U(z)_+$ and $U(z)_- - U(1-z)_+$.

**Results and Discussion**

Water molecules in the SWNT form a single hydrogen-bond chain, and the OH bonds involved in hydrogen bonds are nearly aligned along the nanotube axis and collectively flip their orientations [26, 37, 40]. To quantify the orientation of the water chain, the average angle $\bar{\phi}$ was defined [37], where $\phi$ is the angle between a water molecule dipole and the $z$-axis, and average runs over all the water molecules inside the tube. Most of the time, $\bar{\phi}$ fell between two ranges, $15^0 < \bar{\phi} < 50^0$ and $130^0 < \bar{\phi} < 165^0$, hereafter called + dipole and - dipole states, respectively. The typical cases for + dipole and - dipole states are illustrated in Fig. 1 (A, B). The average duration $T$ for each state was 2.6 ns. For each state with duration longer than 1 ns, data were collected for further analysis.

A total of 152 ns were obtained in the + dipole state from simulation. During those time periods, 1,643 molecules passed from left to right of the SWNT, and 824 molecules passed from right to left, resulting in an average net flux of 5.4 ns$^{-1}$ along the $+z$ axis. The corresponding data were 564 from left to right, 1113 from right to left, and an average net flux of -5.4 ns$^{-1}$, respectively, during 102 ns of the - dipole state.

It is believed that the spontaneous directional motion arising from a thermally fluctuating environment is generally ascribed to the ratchet effect [1-3, 51-58]. The ratchet effect has been successfully applied in man-made devices [51-54] as well as in biological processes [56, 57]. Conventionally, an external asymmetric structure is always required to construct a ratchet potential if there is no external field [1-3, 51-58]. In the present case, however, the SWNT was symmetrical and no external field was applied. The remarkable directional net flux of water molecules results from the asymmetric *potential of mean force* (PMF), which acts as a ratchet potential. In the current study, PMF $U(z)$ at position $z$ was calculated as $U(z) = -k_B T \ln P(z)$ [59], where $P(z)$ is the probability of water molecules appearing at position $z$, and $z$ is the normalized position in the nanotube along the $z$ direction. In Fig. 1(C), the PMFs of water molecules inside the SWNT are shown, denoted by $U(z)_\pm$, for different dipole states. The PMF of the - dipole state has a small systematic displacement along the +z direction compared with that of the + dipole state. The difference between them can be seen more clearly in Fig. 1(C) (bottom). Both $U(z)_+$ and $U(z)_-$ were asymmetric since $U(z)_+ + U(z)_-$ was symmetric. In order to demonstrate that this asymmetry is caused by physical phenomena rather than by the errors raised by a limited simulation timescale, we have shown the data $U(z)_- - U(1-z)_+$ as a red line in the bottom panel of Fig. 1(C). Since the system used here is symmetric, the PMFs of +dipole and -dipole states should have mirror symmetry. That is $U(z)_- - U(1-z)_+ = 0$, if there is no error, which is expected as the simulation time approaches infinity. The value of $U(z)_- - U(1-z)_+$ can be an indication of the error of the PMF due to finite simulation time. It can be seen that the absolute value of $U(z)_- - U(1-z)_+$ is much less than the absolute value of $U(z)_- - U(z)_+$, showing that the numerical error is much less than the asymmetry observed in PMF.

   The asymmetry in PMF comes from the single-file water chain with a concerted dipole orientation. Take the + dipole state for example, with a similar situation for the – dipole state. As shown in Fig. 1(A), the water molecule at the rightmost end inside the channel has a positive dipole along +z direction, so that a hydrogen atom faces the water molecules outside the channel. Since each water molecule inside the channel has a positive dipole along +z direction during a + dipole state, an oxygen atom faces outside at the leftmost end. Due to the large mass difference between the hydrogen and oxygen atoms, it is expected that the water molecules at the leftmost and rightmost ends inside the channel will have different interactions with the outside water molecules, thus these water molecules will have asymmetric equilibrium positions with respect to the center of channel, which influences the density distribution of the whole water chain inside the channel and brings the asymmetry of the PMF along the channel of the + dipole state. According to the method described by Kosztin and Schulten [55], the directions of the net fluxes for the + or - dipole states from $U(z)_+$ and $U(z)_-$ were consistent with simulation results, while the explicit values depended on the model parameters (details can be seen in supplementary material). It is noted that the remarkable unidirectional fluxes in + and - states resulted from the spatial-asymmetrical PMF of the transported water molecules themselves by achieving a ratchet-like effect. The discovery benefited from our analysis of data according to the direction of the concerted hydrogen-bonding orientations.

   The average duration $T$ (=2.6 ns) for each state reaches the timescale of many biomolecular functions [60] such as dissociation of CO, motion of iron, and tilting of heme [61]. This duration is also quite long if compared with the time scale of the component of salvation dynamics of water molecules confined in a biological system, which is around 100-1000 ps [62, 63]. We also note that the average net flux for each state was comparable to the measured 3.9 ± 0.6 ns$^{-1}$ for the biological water channel, aquaporin-1 [40]. Moreover, the average duration can be simply increased in a longer SWNT as shown in Fig. 2. Although the net flux (Fig. 2) decreases for a longer SWNT, the longer duration (together with a larger net flux) is thought to be conceivable in other kinds of nanochannels by modifying factors such as channel-water interaction potentials. In comparison with the value for biological gating, $T$ (=2.6 ns) is still relatively small. However, inside the biological channels, it is possible that the aligned water chain is stabilized by complex interaction with protein residues of the channels such that the duration of each water chain orientation is significantly extended.

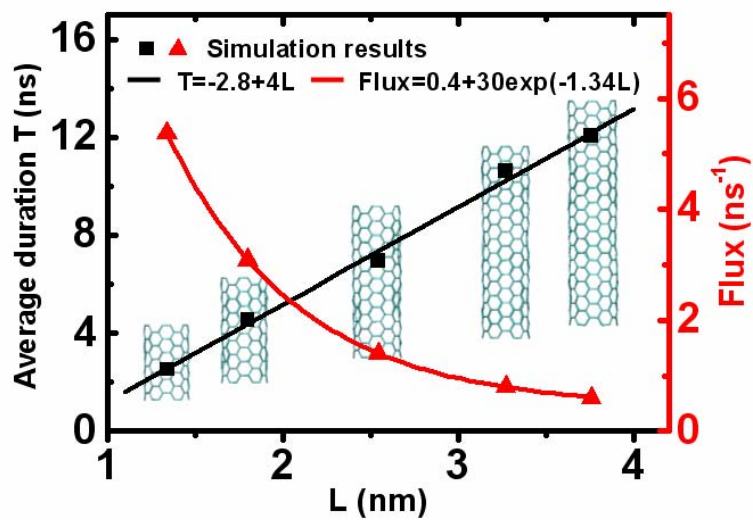

**Figure 2.** The average duration (black) of a hydrogen-bond chain orientation state, and the absolute value of the average net flux (red) in both states with respect to length $L$ of the nanotube. Average duration increases almost linearly, whereas net flux decreases exponentially with respect to $L$.

Since the stable directional transportation of water molecules can maintain 10,000 times the scale of thermal fluctuations (~0.1ps) or longer, useful work may be gained by an elegant external trigger (energy output). It is noted here that such a molecular machine does not violate the second law of thermodynamics. The total duration for both dipole states should be equal and the average net water flux vanishes when the timescale approaches infinity, if there is not any external perturbation to the nanotubes and water molecules. Numerical simulations demonstrate that as the simulation timescale increases, net flux decreases towards zero (see supplementary material). Each + and - state is only the equilibrium (orientation) fluctuation like a gas molecule in a free path. The approach used here extends the timescale of thermal fluctuations from the typical value of $10^{-13}$ s in bulk systems to nanoseconds or longer by one-dimensional confinement and net directed water transportation is achieved within an orientation fluctuation. Thus, the observations in this Letter do not challenge the second law of thermodynamics. A critical finding here is that useful work can be converted from the thermal fluctuation of surroundings in a long time period within an orientation fluctuation, without external input energy. This is an important finding in that once the timescale for such (orientation) fluctuation approaches the timescale of human life, only one or few fluctuations may be observed if there are no external perturbations to the system. Useful work may be gained without input energy for a long enough period of time within this timescale.

To further evaluate the importance of hydrogen bonds of water molecules, the point charges on atoms O and H of the water model were modified so that both number and strength of hydrogen bonds changed, in accordance with previous studies [26, 64]. The average duration for each state was only 0.4 ns and the average net water flux in each dipole state was less than one when the point charges were reduced to 90% of their original values. A stable directed motion of water molecules was not obtained. However, when the point charges increased by 10%, a liquid-gas transition was observed, such as that observed by adjusting the water-carbon interaction [26]. Thus, it appears that water has appropriate point charges on H and O to keep the proper hydrogen bond strength so that both net flux and average duration for each state are large enough.

## Conclusion

Contrary to conventional phenomenon, water molecules can perform self-driven (spontaneous) directed motion spontaneously within a long time period across a nanochannel, without external input energy by achieving a ratchet-like effect. The asymmetric ratchet-like potential solely resulted from the transported water molecules inside the nanochannel that form hydrogen-bonded chains. Importantly, the resulting net water fluxes reach the level of the biological channel. The duration of the spontaneous directed transportation is long (10,000 times over that of thermal fluctuation), reaching the timescale of many biomolecular functions, but limited and it does not violate the second law of thermodynamics.

Since the spontaneous directed transportation is ascribed to the exceptive structure of the water molecules that form hydrogen-bonded chains inside the nanochannel, and given that the generation of the directional movement from thermal fluctuations is critical for the phenomena of biological organization and dynamics, this finding might be helpful in understanding why water is unique [65] and is called life's matrix [66].

### Acknowledgment

The authors thank Prof. Bailin Hao for helpful discussions. This work was supported by Chinese Academy of Sciences, National Science Foundation of China under grants No. 10474109 and 10674146, the National Basic Research Program of China under grant Nos. 2006CB933000 and 2006CB708612 and Shanghai Supercomputer Center of China.

**Table of contents of Graphic**

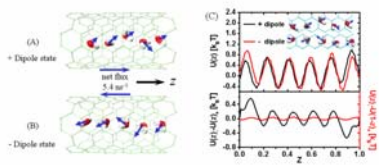